\documentclass{appolb}
\usepackage{graphicx}

\begin{document}
\title{Transformation between complex scattering length and binding
energy\footnote{Contributed paper at the II International
Symposium on Mesic Nuclei, held 22-24 September, 2013 at Krak\'ow, Poland.
All correspondences should be sent to haider@fordham.edu} \\}
\author{ Q. Haider
\address{Department of Physics and Engineering Physics\\ Fordham University\\
Bronx, N. Y. 10458, U.S.A.}
\and
 Lon-chang Liu
\address{Theoretical Division\\ Los Alamos National Laboratory\\
Los Alamos, N. M. 87545, U.S.A.}}

\maketitle

\begin{abstract}

The use of scattering length of particle-target interaction due to real-valued potential
to study the bound states of the particle-target system is well known in nuclear and atomic
physics. In view of the current interest in using $\eta$-nucleus scattering length to infer
the existence of $\eta$-mesic nucleus, we derive general analytic expressions that relate
the binding energy and half-width of an unstable bound state to the complex-valued scattering
length due to the same particle-target interaction.

\end{abstract}

\pagebreak

\section{Introduction} \label{sec1}

The existence of eta-mesic nucleus, a novel nucleus with an excitation energy of about 540 MeV,
first predicted more than 25 years ago \cite{hai1}, has led to several theoretical studies \cite{hai2, kelk} aimed
at understanding the underlying interaction between an eta ($\eta$) meson and a nucleus.
While there is difference of opinion on the
size of the nucleus in which the $\eta$ can be bound, there is general agreement that theoretical calculations,
irrespective of the formalism and $\eta$N interaction model used, provide compelling reason to believe that $\eta$-mesic
nucleus does indeed exist with nuclear mass number greater than 10 \cite{hai2}.

Since 1986, the search for $\eta$-mesic nucleus has been conducted at various laboratories in the U.S.A., Europe and Japan.
The most promising indication of the existence of $\eta$-mesic nucleus
comes from an experiment performed at COSY-GEM (J\"ulich)~\cite{exp3} using the (p,$^{3}$He) reaction on $^{27}$Al.
This experimental result has in turn further encouraged efforts in searching the possibility of forming $\eta$-mesic nucleus
in light nuclei such as $^{3,4}$He \cite{kelk}-\cite{krzm}, as advocated by some researchers.
Clearly, this latter possibility is directly related to the magnitude of
the $\eta$-nucleus scattering length $a$.

In a recent work, Niskanen and Machner \cite {nis2} explored the relation between the binding energy and width of
an $\eta$-mesic nucleus and the complex-valued $\eta$-nucleus scattering length. One of the underlying ideas of such
study is that an experimental
determination of $a$ via final-state interaction between $\eta$ and a nucleus  would indicate the existence or
nonexistence of an $\eta$-nucleus $s$-wave bound state. Independent of the way how the scattering length or the binding
energy is modeled or measured, we will see that the functional relation between these two observables is in fact
model-independent. Hence, this dependence is an interesting subject on its own. In this paper, we present
the model-independent features of
this functional dependence.

\section{Polar representation of observables} \label{sec2a}

The $s$-wave scattering amplitude is given by

\begin{equation}
f=\frac{e^{2i\delta}-1}{2ik} = \frac{1}{k\cot\delta - ik},
\label{1}
\end{equation}

\noindent
where $\delta$ is the phase shift which is complex-valued when an optical potential is used in
calculating it and $k$ is the final-state channel momentum. For exponentially bound potentials,

\begin{equation}
k\cot\delta = \frac{1}{a} + \frac{1}{2}rk^{2}+ ...,
\label{2}
\end{equation}

\noindent
where $a$ is the scattering length. In the limit $k\rightarrow 0$, the first term dominates, i.e.,

\begin{equation}
k\cot\delta = \frac{1}{a}
\label{3},
\end{equation}

\noindent
and

\begin{equation}
f=\frac{1}{1/a - ik}.
\label{4}
\end{equation}

\noindent
The bound-state pole occurs at $k_{pol}=-i/a$ which leads to

\begin{equation}
k_{pol}^{2}=-\frac{1}{a^{2}}.
\label{5}
\end{equation}

It is useful to use the polar representation in the complex $a$ plane so that

\begin{equation}
a=| a| \exp (i\gamma ) \equiv x+iy,
\label{6}
\end{equation}

\noindent
with

\begin{equation}
x = \mbox{Re} [a] =| a| \cos\gamma ,\;\;
y = \mbox{Im} [a] =| a| \sin\gamma ,\;\;
\gamma= \arctan \left (\frac{y}{x}\right ).
\label{7}
\end{equation}

\noindent
It follows that

\begin{equation}
 k_{pol}^{2}=-\frac{1}{| a|^{2}}\;\exp (-2i\gamma) \ .
\label{8}
\end{equation}

\noindent
The complex energy $B$ is, therefore, given by

\begin{equation}
B=\frac{k_{pol}^2}{2\mu} =  -|B|\ \exp (-2i\gamma),
\label{10}
\end{equation}

\noindent
where $\mu$ is the reduced mass of the bound particle and

\begin{equation}
 |B| = \frac{1}{2\mu|a|^{2}} \ .
\label{10b}
\end{equation}

\noindent
In the Cartesian representation,

\begin{equation}
 B \equiv E-i\frac{\Gamma}{2} \equiv  u + iv\ ,
\label{9}
\end{equation}

\noindent
where $E$ and $\Gamma/2$ denote, respectively, the binding energy and half-width of the bound state. It
follows from Eqs.(\ref{6}) and (\ref{7}) that

\begin{equation}
 u \equiv E\ = \ -\frac{x^2-y^2}{2\mu|a|^4}  ,
\label{11}
\end{equation}

\begin{equation}
 v \equiv - \frac{\Gamma}{2}\ =  \frac{2xy}{2\mu|a|^4} \ .
\label{12}
\end{equation}

The polar representation in the $a$-plane, Eq.(\ref{6}), was used by Balakrishnan et al.~\cite{bala}
to calculate the energies and widths of bound states as functions
of complex scattering lengths in multi-channel atom-molecule collisions.
In section \ref{sec2} we expand the polar representation to include the $k_{pol}$- and $B$-planes, and
determine the respective physical domains. Using these
domains, we solve in section \ref{sec3} the inverse problem, namely,
finding the value of complex $a$ for a given value of complex energy $B$.

\section{Physical domains on the complex planes} \label{sec2}

If we denote
\begin{equation}
k_{pol} =\ R\ +\ iI \ ,
\label{13}
\end{equation}

\noindent
then

\begin{equation}
k_{pol}^{2}=R^{2}-I^{2}+2iRI.
\label{14}
\end{equation}

\noindent
Hence, we obtain from Eqs.(\ref{10}) and (\ref{9})

\begin{equation}
2\mu E= \ R^{2}-I^{2}\ = - \frac{1}{|a|^{2}}\;\cos 2\gamma ,
\label{15}
\end{equation}

\begin{equation}
2\mu \left ( \frac{\Gamma}{2}\right )= - 2RI = - \frac{1}{|a|^{2}}\;\sin 2\gamma .
\label{16}
\end{equation}

\noindent
Because $E<0$ (the bound state) and $\Gamma >0$ (by definition),
Eqs.(\ref{15}) and (\ref{16}) are, respectively, equivalent to

\begin{equation}
(R^{2}-I^{2}) <0 \;\; \mbox{and} \;\; RI <0.
\label{17}
\end{equation}

The first inequality requires $|R|<|I|.$ Since a bound state has $I>0$ (i.e. a decaying outgoing wave),
the second inequality then requires $R<0.$ In summary, the requirements $|R|<|I|,\;\;I>0$ and $R<0$
indicate that the bound-state (in the literature it is also called quasi-bound state) poles are situated in
the second quadrant (but above the diagonal line) of the complex $k_{pol}$-plane.

Equations~(\ref{15}) and (\ref{16}) further indicate that $E<0$ and $\Gamma >0$ require, respectively, that
$\cos 2\gamma > 0$ and $\sin 2\gamma <0.$ This in turn requires $3\pi /2 < 2\gamma < 2\pi$ or
$3\pi /4 < \gamma <\pi$. The complex scattering length, $a$, is therefore situated in the second quadrant
but below the diagonal line in the complex $a$-plane.
In other words, on the $a$-plane the scattering length satisfies simultaneously

\begin{equation}
\mbox{Im}[a]>0,\;\;\mbox{Re}[a]<0,\;\;|\mbox{Im}[a]|<|\mbox{Re}[a]|.
\label{18}
\end{equation}

\noindent
The third inequality was also shown in Ref.\cite{hai2}. The physical domains in the $a$-,
$k_{pol}$-, and $B$-planes are shown in Figs.1-3, respectively.
When the polar angle, $\gamma$, in the $a$-plane turns counter clockwise, the corresponding polar
angles in the $k_{pol}$- and $B$-planes turns in the opposite direction (see section ~\ref{sec4.1}).

\begin{figure}
\includegraphics[width=1\columnwidth]{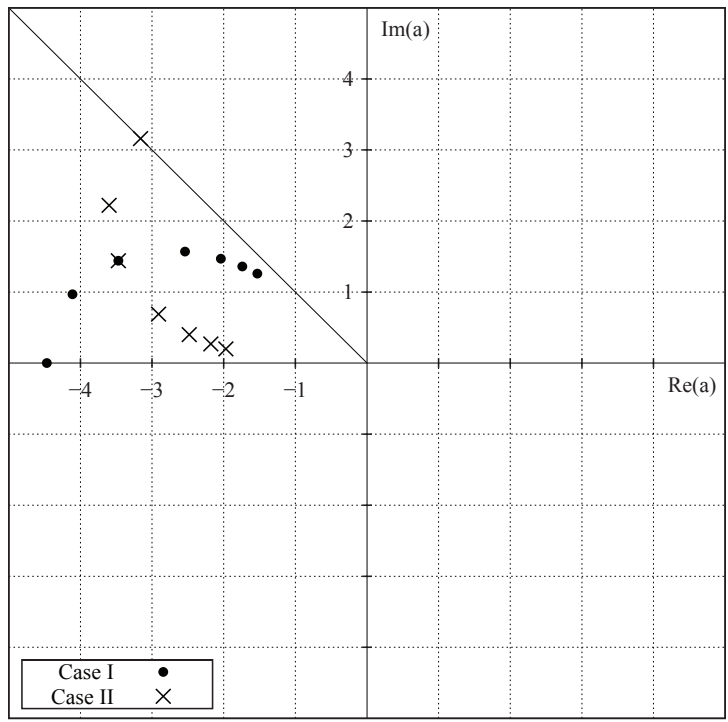}
\caption{The complex scattering length plane. The physical domain is the entire lower
triangular region of the 2nd quadrant. The meaning of the solid circles and the crosses
are given in the text.}
\label{Fig1}
\end{figure}

\begin{figure}
\includegraphics[width=1\columnwidth]{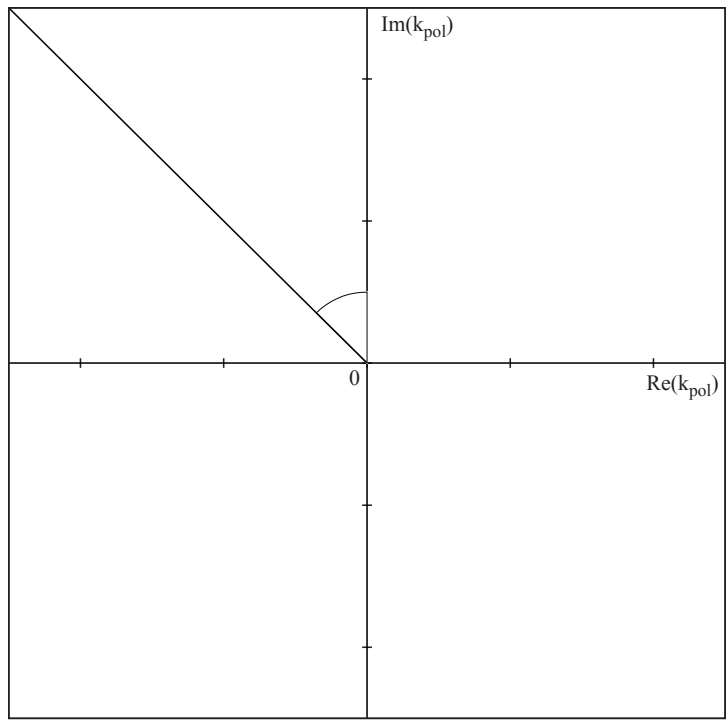}
\caption{The complex momentum plane. The physical domain
is the entire upper triangular region of the 2nd quadrant. The arc near the origin indicates the
corresponding polar-angle range.}
\label{Fig2}
\end{figure}

\begin{figure}
\includegraphics[width=1\columnwidth]{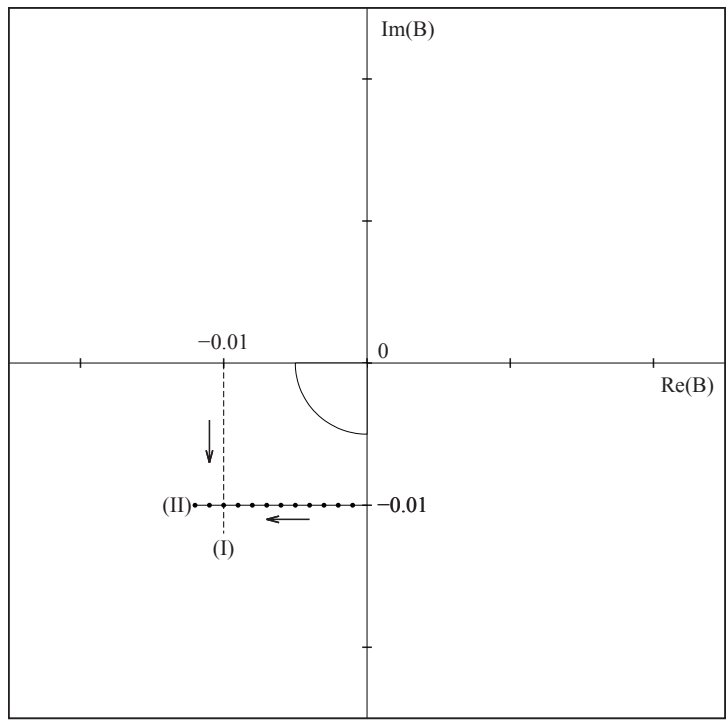}
\caption{The binding-energy plane. The physical domain
is the entire 3rd quadrant. The arc indicates the polar-angle range.
The trajectories shown by the downward dashed line and the
horizontal linked-dotted line are explained in the text.}
\label{Fig3}
\end{figure}

\section{Inverse mapping}\label{sec3}

From Eqs.(\ref{10b}), (\ref{11}), and (\ref{12}), we get

\begin{equation}
 u = -2\mu|B|^2(x^2-y^2) \ ,
\label{19}
\end{equation}

\noindent
and

\begin{equation}
 v = 2\mu|B|^2 (2xy) \ .
\label{20}
\end{equation}

The inverse mapping is obtained by solving the above coupled equations and the result is:

\begin{equation}
  x = \left (-\frac{1}{2|B|\sqrt{\mu}} \right )\sqrt{ -u + \sqrt{u^2 + v^2} } \ ,
\label{23}
\end{equation}

\begin{equation}
 y = \left (\frac{1}{2|B|\sqrt{\mu}} \right )\sqrt{ u + \sqrt{u^2 + v^2}} \ .
\label{24}
\end{equation}

\noindent
In choosing the branch of the square roots, the properties associated with the
physical domains discussed in section~\ref{sec2} have been used.

\section{Results and discussion}\label{sec4}

\subsection{Application of the inverse mapping}\label{sec4.1}

As examples of the application of the inverse mapping,
we have considered two cases. The departing trajectories on the $B$-plane are shown in Fig.3.
Case I corresponds to fixing $E$ but varying $\Gamma /2$ in the direction given
by the downward arrow along the dashed line. Case II corresponds to fixing
the value of $\Gamma/2$ while varying $E$ along the horizontal linked-dotted line in
the direction shown by the leftward arrow. The resulting $a$ (calculated with $2\mu = 5~\mbox{fm}^{-1}$)
are shown, respectively, as the left-to-right solid circles and
the descending crosses in Fig.1. One notes that the polar angles,
$\gamma$, of the successive left-to-right solid circles in Fig.1 turn clockwise while those of the
original $B$-points along the downward dashed line in Fig.3
turn counter clockwise. This reversal of the sense of the
turnung is a consequence of the opposite signs in front of the polar angles in Eqs.(\ref{6}) and (\ref{10}).

\subsection{Role of nuclear mass in the binding of $\eta$}\label{sec4.3}

From Eqs.(\ref{23}) and (\ref{24}), one notes
that the magnitudes of the real and imaginary parts of the scattering length $a$ are inversely proportional to
$\sqrt{\mu}$. For $\eta$-mesic nucleus, $\mu$ is the reduced mass of $\eta$, which increases
as nuclear mass increases. This in turn indicates that for $\eta$ to have a given binding energy $|E|$
in a lighter nucleus, it requires a larger $|\mbox{Re}[a]|$ than that for $\eta$ to have the same
binding energy $|E|$ in heavier nuclei. This is why the $\eta$N model of Ref.\cite{hai1} does not
predict the existence of $^{3,4}$He$_\eta$ while models giving larger $|a_{\eta N}|$ (and,
hence, larger $|a_{\eta A}|$\ ) do predict these light $\eta$-mesic nuclei.

\subsection{Remark on the sign convention of the scattering length}\label{sec4.4}

The scattering length approximation is often written as

\begin{equation}
k\cot\delta = -\frac{1}{a}
\label{25}
\end{equation}

\noindent
To avoid confusion, let us denote $- a={\cal A}$ in Eq.(\ref{19}).
The polar representation of $a$, Eq.(\ref{6}), then leads to the following equations:

\begin{equation}
{\cal A}=-|a|e^{i\gamma}= |a|e^{i(\pi +\gamma )}=|a|e^{i\theta},
\label{26}
\end{equation}

\noindent
where we have defined $\theta = \pi +\gamma$. By repeating the algebra leading Eq.(\ref{3})
to Eq.(\ref{18}), we have noted that the forms of Eqs.(\ref{8}), (\ref{10}), (\ref{15}), (~\ref{16}) remain unchanged,
except that in these equations the variable $2\gamma$ will be replaced by $2\theta$. However, this
variable change has no consequence on the final results because $\cos 2\theta = \cos 2\gamma$ and
$\sin 2\theta = \sin 2\gamma$. In other words, when one goes from $(\mbox{Re}[a],\mbox{Im}[a])$ to
determine $(E,\Gamma /2)$, the result is independent of the sign of the scattering length, $a$.
This is because $k_{pol}^{2}$ is independent of the sign of $a$. However, because $- a = {\cal A}$, the physical
domain on the ${\cal A}$-plane will be the upper triangular region
 of the the 4th quadrant. If we define ${\cal A} = x' + iy'$, then
$x'=-x$ and $y'=-y$, with the $x$ and $y$ given, respectively, by Eqs.(\ref{23}) and (\ref{24}).
However, we will still have the inequality $|\mbox{Im}[{\cal A}]|<|\mbox{Re}[{\cal A}]|$, the same as the
third inequality in Eq.(\ref{18}).

\section{Summary}\label{sec5}

We have studied the complex mappings  $a \rightarrow B$ and $B\rightarrow a$.
By using the physics implied by the momentum of a
bound state, we have determined the physical domain of $a$ (Fig.1).
The $a$ satisfies the properties given in Eq.(\ref{18}).

Our analytic expressions (Eqs.(\ref{23}) and (\ref{24})) are interaction-model independent so long as
the potential of the particle-target interaction belongs to the class of exponentially bound potentials so
that the low-energy expansion, Eq.(\ref{2}), can be made. The only kinematic approximation used in our derivation
is $k_{pol}^2/2\mu \simeq \sqrt{k_{pol}^2 + \mu^2} - \mu$ which is a very good approximation for
binding problems. We emphasize that the model depedence of the interaction dynamics does come into play
when one theoretically calculates the scattering length, $a$, and the binding energies, $B$.
However, once one of these two observables is calculated (or measured), then the remaining one is determined
by Eqs.(\ref{15}) and (\ref{16}) or, vice versa, by Eqs.(\ref{23}) and (\ref{24}).
These analytic expressions offer, therefore, a means of making consistency test of interaction models or measurements.
We mention, among others, that these analytical expressions can be calculated readily by means of hand calculators.
Finally, they can be used to gain insight into the trend of bound-state formation as elucidated by the example
discussed in section ~\ref{sec4.3}.

\section{Ackknowledgement}\label{ack}
We would like to thank Ariel Fragale, a senior physics major at Fordham University for her assistance with
the figures. One of us (Q.H.) would like to thank Professor Pawel Moskal of Jagiellionian University for
the hospitality extended to him during his stay at Krak\'ow.

\end{document}